\documentclass{ws-mpla}
\begin{document}

\markboth{Hanqing Zheng} {Low lying scalar resonances and chiral
symmetry}

\catchline{}{}{}{}{}

\title{LOW LYING SCALAR RESONANCES AND CHIRAL SYMMETRY\footnote{Invited plenary talk given at
CHIRAL 2007, Nov. 13-16, 2007, RCNP, Osaka, Japan.}}

\author{\footnotesize HANQING ZHENG }

\address{Department of Physics, Peking University, Beijing 100871,
China\\ e-mail: zhenghq@pku.edu.cn }

\maketitle


\begin{abstract}
Current theoretical studies on the $\sigma$ and $\kappa$ resonances
are reviewed. It is emphasized that all evidences accumulated so far
are consistent with the picture that the $\sigma$ meson is the
chiral partner of the Nambu--Goldstone bosons in a linear
realization of chiral symmetry.

\end{abstract}

\ccode{PACS Nos.: 12.30.Rd, 11.55.Bq, 12.39.Fe}

\section{The Renaissance of the $\sigma$ Meson}

\subsection{Early studies on the physics related to the  $\sigma$ meson}
The $\sigma$ particle was firstly introduced by Gell-Mann and Levy
in association with the linear $\sigma$  model.\cite{gellmann} In
such a model, the $\sigma$ meson develops a non-vanishing vacuum
expectation value which triggers the spontaneous breaking of chiral
symmetry. Pions act as the (pseudo-)goldstone bosons associated with
the spontaneous chiral symmetry breaking (S$\chi$SB). In the linear
$\sigma$ model the $\sigma$ meson fills in the linear chiral
multiplet together with pions. The $\sigma$   and the   pion fields
transform and mix with each other under chiral rotations, and before
S$\chi$SB the $\sigma$ field and the $\pi$ field were essentially
the same.

What is  described above only repeats standard text book content.
The question is then whether the theory can accomodate for
experimental data. Early studies on nuclear physics require $\sigma$
in order to cancel the large $\pi$N scattering lengths caused by the
$\pi$ nucleon Born term.~\cite{robilotta} However these studies are
based on models with linearly realized chiral symmetry. In such
models, there must exist a large cancelation between the $\sigma$
contribution and the $\pi$ contribution at low energies in order to
obey various constraints from soft pion theorems. One example of the
latter  is the Adler zero condition.

Non-linear realization of chiral symmetry was later
discovered,~\cite{coleman} which satisfies all low energy theorems
induced by PCAC.~\cite{PCAC}  Hence it is suggested that the
$\sigma$ is not necessary for chiral symmetry breaking. Chiral
perturbation $theory$ ($\chi$PT) is established based on the
non-linear realization of chiral symmetry.~\cite{weinberg,GL84} It
is a model independent approach and successfully describes strong
interaction physics at (very) low energies. Furthermore it was shown
that the (renormalizable) linear $\sigma$ model is not QCD at low
energies.~\cite{GL84}

On the other side, remarkable efforts have been made by  physicists
who insist on the existence of $\sigma$ meson, which  led to the
return of the $\sigma$ meson in PDG after disappearing for more than
30 years.\cite{tornq} Most of these studies and hence conclusions
are model dependent, but are however   important in keeping the
thoughts on the right trajectory. On experimental side, it is also
very difficult to claim the discovery of the $\sigma$ pole from
experimental data, since the $\sigma$ meson, if exists, has to be a
very broad resonance. In other words, $s$ wave I=0 channel $\pi\pi$
interaction at low energies is of highly non-perturbative, strong
interaction nature. It is not easy to extract the $\sigma$ pole from
the background contributions.\footnote{The major difficulties in
accepting the $\sigma$ resonance, and how can they be overcome, have
been reviewed in Ref.\cite{mont}.} As a consequence, previous
conclusions in supporting the existence of the $\sigma$ meson are
not very convincing.

\subsection{The $\sigma$ meson must exist to adjust $\chi$PT to experiments}

The situation still seemed to be rather confusing: chiral
perturbation theory does not seem to support the existence of the
$\sigma$ meson. On the other hand, see Fig.~1, the steady rise of
the I,J=0,0 channel $\pi\pi$ scattering phase shift data below 1GeV,
provided by the old CERN-Munich collaboration, left  a big puzzle
for imagination.
\begin{figure}[h]
\begin{center}
\mbox{\epsfxsize=75mm \epsffile{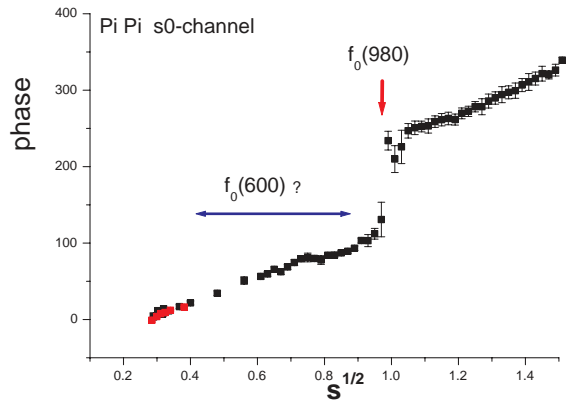}}
\caption{\label{00phaseshift}IJ=00 channel $\pi\pi$ scattering phase
shift data from Ref.~\protect\cite{munich1}
 and
Ref.~\protect\cite{E865}.
}
\end{center}
\end{figure}

In my personal opinion, the clearest way to reveal the very
existence of the $\sigma$ meson hidden behind the phase shift data
in Fig.~\ref{00phaseshift} may be to write down a dispersion
relation for the $\sin(2\delta_0^0)$.\cite{XZ00} Define
 \begin{equation} F(s)\equiv \frac{1}{2i\rho}(S(s)-{1\over S(s)}) \
 ,
 \end{equation}
 it is easy to understand that $F(s)$ is analytic across the elastic
 cut of $\pi\pi$ elastic scattering and
\begin{equation}
 \sin(2\delta_\pi)=\rho F \ ,
 \end{equation}
 where $\rho=\sqrt{\frac{s-4m_\pi^2}{s}}$. One assumes a high energy Regge behavior for the
 partial wave amplitude, hence $F(s)$ satisfies a once-subtracted dispersion relation,\cite{XZ00}
\begin{eqnarray}\label{Fss}
  F(s)&=&
 \alpha+ \sum\mbox{poles}+{1\over\pi}\int_L{{\rm Im}_LF(s')
  \over s'-s} ds'
  +{1\over\pi}\int_{R}{{\rm Im}_RF(s')
  \over s'-s} ds'\ ,
   \end{eqnarray}
where $\alpha$ is a subtraction constant and the rest contribution
to $F(s)$ come either from poles or cuts.
  For $\pi \pi$ scatterings $L=(-\infty,0]$ and
 R starts from first inelastic threshold, practically $R=[ 4m_K^2,\infty)$.

The right hand cut can be estimated using experimental data and it
is found that its contribution is not important at low energies. The
left hand cut is estimated using $\chi$PT. As seen in Fig.~2, the
left hand cut contribution to the phase shift is negative and
concave whereas the experimental data curve on $\sin(2\delta_0^0)$
is positive and convex. Comparing with Fig.~1, Fig.~2 clearly
demonstrates that it is necessary to include the $\sigma$ meson to
adjust chiral perturbation theory to experiments, according to
Eq.~(\ref{Fss}).
\begin{figure}[h]
\begin{center}
\mbox{\epsfxsize=85mm\epsffile{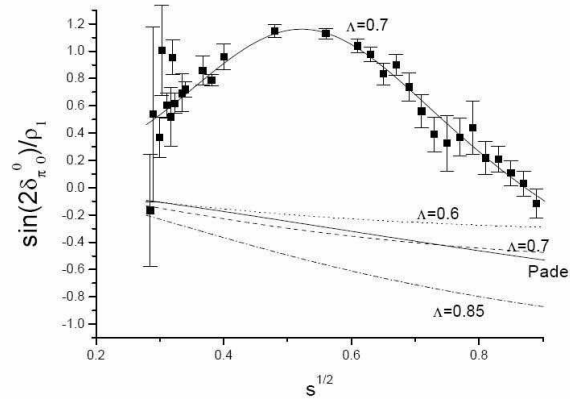}}
\caption{\label{fig1}Experimental data and fit curve of
$\sin(2\delta_0^0)$. Various estimates on left hand cut using
$\chi$PT and its Pad\'e approximant are also drawn. }
\end{center}
\end{figure}

Evidences for the existence of $\sigma$ and $\kappa$ resonances are
also reported in recent years production
experiments~\cite{sigma-exp}-\cite{kappa-exp}.

\section{Dispersive Analyses to the $\sigma$ Meson}
Because the physics related to the $\sigma$ meson and the $\kappa$
meson in $s$ wave I=1/2 channel $\pi K$ scatterings are of highly
non-perturbative nature, a dispersive analysis is hence needed to
extract the physical information in a model independent way. In the
next we briefly review a new dispersive approach we have developed
which is particularly suitable to study low lying resonance
structure.
\subsection{Factorized $S$ matrix element and separable singularities
-- the PKU parametrization form } One constructs, using analyticity,
unitarity, and $partial$ $wave$ dispersion relations, a factorized
form for elastic scattering $S$ matrix
element:\cite{zheng01}--\cite{zhou06}
\begin{equation}\label{Separable}
 S^{phy.}=\prod_iS^{R_i}\cdot S^{cut}\ ,
\end{equation}
where $S^{R_i}$ denotes the $i$-th $second$ sheet pole contribution
and $S^{cut}$ denotes the contribution from cuts or background. The
information from higher sheet poles is hidden in the right hand
integral which consists of one part of the total background
contribution,
\begin{eqnarray}\label{fs}
 S^{cut}&=&e^{2i\rho f(s)}\ ,\nonumber\\
 f(s)&=&\frac{s}{\pi}\int_{L}\frac{{\rm
 Im}_Lf(s')}{s'(s'-s)}+\frac{s}{\pi}\int_{R}\frac{{\rm
 Im}_{R}f(s')}{s'(s'-s)}\ .
\end{eqnarray}
The `left hand' cut $L=(-\infty, 0]$ for equal mass scatterings and
may contain a rather complicated structure for unequal mass
scatterings. The right hand cut $R$ starts from first inelastic
threshold to positive infinity. It can be demonstrated\cite{zhou06}
that the dispersive representation for $f$ is free from the
subtraction constant. Again the right hand cut can be  estimated
from experimental input. Nearby left hand cut ($s>-32m_\pi^2$ for
$\pi\pi$ scatterings if assuming Mandelstam representation) can in
principle be estimated from experimental data as well, using
Froissart--Gribov projection formula. Nevertheless, one expects such
an estimate gives more or less the same effect as using $\chi$PT
results when estimating the nearby left cut. There is no reliable
way to estimate further left hand cut effects, nevertheless physics
around physical threshold should not be strongly affected by what
might happen far away in the complex $s$ plane. Moreover, further
left cut contribution as defined by Eq.~(\ref{fs}) is very mild
which is understood by the fact that the integrand appeared in
Eq.~(\ref{fs}) is of logarithmic form. Numerical evaluation
justifies this argument.\cite{zhou05}-\cite{zhou06}

Estimates in various channels of $\pi\pi$ and $\pi K$ scatterings
reveal a common feature: all the background contributions as defined
in Eq.~(\ref{fs}) are numerically found to be negative! This fact is
actually crucial to establish the existence of the $\sigma$ and
$\kappa$ pole in the present approach and also helps greatly in
stabilizing the pole location in the data fit. It is interesting to
notice that, there actually exists a correspondence of
Eq.~(\ref{Separable}) in quantum mechanical scattering theory,
obtained sixty years ago:\cite{NingHu}
 \begin{equation}\label{NHu}
S(k)=e^{-2ikR}\prod^{\infty}_{1}\frac{k_n+k}{k_n-k}\ ,
\end{equation}
 where $k$ is the (single) channel momentum and $k_n$ pole locations in the
 complex $k$ plane. The above formula is written down for any finite range
 potential, in $s$ wave. It amazing to notice that the
 Eq.~(\ref{NHu}) automatically predicts a negative background
 contribution!

\subsection{The $\sigma$ and $\kappa$ pole locations}
The Eq.~(\ref{Separable}) can actually be understood as a simple
combination of single channel unitarity and the partial wave
dispersion relation.\cite{juanjo07} In the data fit it is found that
the parametrization form is sensitive to $S$ matrix poles not too
far away from physical threshold, hence providing a useful tool to
explore the broad resonance $\sigma$ and $\kappa$.

In the data fit it is found that crossing symmetry also plays an
important role in fixing the $\sigma$ pole location. Taking this
fact into account\cite{zhou05} it gives the $\sigma$ pole location
at $ M_\sigma=470\pm 50\mbox{MeV}\ ,\,\,\, \Gamma_\sigma=570\pm
50\mbox{MeV}\ , $ in good agreement with the determination using
more sophisticated Roy equation analysis.\cite{caprini} The
application of Eq.~(\ref{Separable}) to LASS data\cite{LASS} also
unambiguously establish the existence of the $\kappa$ meson with the
pole location:\cite{zhou06} $ M_\kappa=694\pm 53\mbox{MeV}\ ,\,\,\,
\Gamma_\kappa=606\pm 59\mbox{MeV}\ , $ which are also in agreement
with the  later determination on $\kappa$ pole parameters using
Roy--Steiner equations.\cite{descotes}

\section{The physical properties of  $\sigma$ and $\kappa$ }
Though the existence of the broad $\sigma$ and $\kappa$ resonances
are firmly established, its nature remains somewhat mysterious.
Especially the question remains completely open on how to understand
it from the underlining theory, QCD. Though it is natural to attempt
to relate  $f_0(600)$ or the $\sigma$ meson to the quantum
excitation of the order parameter $<\bar \psi\psi>$, a proof at the
fundamental level is still missing. The first thing to notice is
that the similarity between the $\sigma$ and $\kappa$ excludes very
likely the possibility that the $\sigma$ contains a large gluball
component. Efforts have been made in trying to understand the
$\sigma$ meson using lattice QCD technique,\cite{KFLiu} but the
approach is still at the stage of prematurity.

Most  papers devoted to the study of physical interpretation of the
$\sigma$ or $f_0(600)$ meson  are at phenomenological level, using
for example, linear sigma model at hadron level,\cite{Lsigma} at
quark level\cite{Lsigmaq}, or the ENJL model\cite{ENJL}. Also the
$\sigma$ meson are considered as a tetra quark state,\cite{jaffe} as
a dynamically generated resonance from chiral perturbation theory
lagrangian,\cite{pelaez} or as a resonance generated from 3P$_0$
potential model.\cite{eef}

It is of course very helpful, when trying to understand $\sigma$, to
investigate the $\kappa$, $f_0(980)$, $a_0(980)$ simultaneously. One
of the most challenging problem is to understand not only the mass
spectrum but also the widely spread widths between these lightest
scalars in different channels. This is recently reinvestigated using
the ENJL model.\cite{su} It is found that the masses and widths of
these lightest scalars, except the $f_0(980)$, can be understood
simultaneously, taking them as the chiral partners of the $SU(3)$
pseudo-goldstone bosons, within linearly realized chiral
symmetry.\cite{su} The estimate is very crude because it is based
upon a $K$-matrix unitarization approach. But one may hope the study
may be of help to grasp the major physics and the qualitative
picture presented by  these lightest scalars.

An examination to the unitarized chiral perturbative amplitudes
finds a light and broad pole on the complex $s$ plane, in the
I,J=0,0 channel of $\pi\pi$ scatterings. It is also found that the
$N_c$ trajectory of the $\sigma$ meson has a non-typical behavior as
comparing with that of a normal resonance, e.g., a $\rho$ pole.
Hence it is argued that the $\sigma$ is a dynamically generated
resonance from a lagrangian without the $\sigma$ degree of
freedom.\cite{pelaez}. This idea has been carefully
examined\cite{sun,guo}, and it is found that the [1,1] Pad\'e
approximation leads to a `$\sigma$' pole falling back to the real
$s$ axis in the large $N_c$ limit.  A correct understanding to what
does the [1,1] Pad\'e approximant  mean is obtained through these
studies. It is also pointed out that the bent structure of the
$\sigma$ pole trajectory with respect to $N_c$ found in [1,1] Pad\'e
approximant is in qualitative agreement with what one finds in
$O(N)$ $\sigma$ model, hence suggesting a fundamental  role of the
light and broad resonance pole played at lagrangian level, even
though it can be generated from certain dynamical
approximations.\cite{guo}

 On the other hand,  both lattice
simulation\cite{KFLiu} or QCD sum rule (SR) approach\cite{QCDsr}
seem to suggest  that the `$\sigma$' pole couples strongly to quark
quadru-linear currents. Based on this observation, one may argue
that the $\sigma$ is a tetra quark state. Nevertheless at low
energies quark and gluons are not the appropriate degrees of freedom
to interpret QCD. To understand this, let us ask a question whether
a pion is a $\bar qq$ state. We know that $\pi$ is a
pseudo-goldstone boson, i.e., a collective excitation of QCD. Hence
in the chiral limit, the pion wave function in the Fock space
expansion contains equally important $\bar qq$, $\bar q^2q^2$, $\bar
q^3q^3$, $\cdots$ components. Then can one conclude that $\pi$ is
not a $\bar qq$ state, even if it contains $\bar qq$ as the leading
component providing the quantum number? If the $\sigma$ meson is the
chiral partner of the $\pi$ field in a lagrangian with a linear
realization of chiral symmetry, it must share many properties of the
pion field, i.e, being also a collective excitation. The lattice or
QCD SR results, in my opinion, seem to support this simple picture
that the $\sigma$ is indeed the chiral partner of the $\pi$ fields.
The leading component in the $\sigma$ wave function is still $\bar
qq$ with vacuum quantum number. Actually the clearest way to see the
$\bar q q$ component is in the large $N_c$ limit, where $\bar q q$
becomes dominant, otherwise the amplitude will have difficulty to
fulfil the requirements of crossing symmetry and analyticity.
\cite{sun,juanjo07} Therefore a calculation in lattice or QCD SR
with a $\bar qq$ component would be more realistic and
welcome.\cite{QCDsr2}

Furthermore, a quark quadru-linear current is actually not
distinguishable from a product of two quark bi-linear current. In
this sense it is also ambiguous to state that the $\sigma$ meson is
a tetra quark state even if there is a large $\bar q^2q^2$ component
inside the $\sigma$. The only un-ambiguous way to explore the tetra
quark state is through searching for states with exotic quantum
numbers. However, it is shown that such states in general are at
least not favorable to be formed.\cite{jido}

To summarize,  the results obtained so far from theoretical studies
to the $\sigma$ meson are well consistent with the picture that it
is the chiral partner of the pesudo-goldstone bosons of QCD. The
$\sigma$ meson contains a $\bar qq$ seed, but is heavily
renormalized by $\pi\pi$ continuum. The latter shifts its pole from
near real axis to a place deep in the $s$ plane.

\section*{Acknowledgments}

I would like to that the organizers of CHIRAL 2007, especially
Profs. H.~Toki, A.~Hosaka and T.~Kunihiro for their kind
hospitality. This work is supported in part by National Natural
Science Foundation of China under contract number
 10575002, 
and  10721063.

\end{document}